\documentclass[conference]{IEEEtran}
\IEEEoverridecommandlockouts
\usepackage{cite}
\usepackage{amsmath,amsfonts}
\usepackage{graphicx}

\usepackage{textcomp}
\usepackage{xcolor}
\usepackage{CJKutf8}
\usepackage{tabularray}
\usepackage{subfigure}
\usepackage{caption}
\usepackage{multirow}
\usepackage{bm}
\usepackage{amsmath,amsfonts}
\usepackage{algorithm}
\usepackage{array}
\usepackage[caption=false,font=normalsize,labelfont=sf,textfont=sf]{subfig}
\usepackage{textcomp}
\usepackage{stfloats}
\usepackage{url}
\usepackage{verbatim}
\usepackage{graphicx}
\usepackage{cite}
\usepackage{amsmath,amsfonts}
\usepackage{array}
\usepackage[caption=false,font=normalsize,labelfont=sf,textfont=sf]{subfig}
\usepackage{textcomp}
\usepackage{stfloats}
\usepackage{url}
\usepackage{verbatim}
\usepackage{graphicx}
\def\BibTeX{{\rm B\kern-.05em{\sc i\kern-.025em b}\kern-.08em
    T\kern-.1667em\lower.7ex\hbox{E}\kern-.125emX}}
\usepackage{balance}
\usepackage{amsmath}
\usepackage{makecell}
\usepackage{array}
\newcolumntype{C}{>{\centering\arraybackslash}X}
\usepackage{epstopdf}

\usepackage{changes}
\usepackage{textcomp}
\usepackage{xcolor}
\usepackage{tabularx,booktabs}
\usepackage[flushleft]{threeparttable}
\usepackage{booktabs,ragged2e}
\usepackage{multirow}
\usepackage{subcaption}
\usepackage{bm}

\usepackage{amsthm}

\allowdisplaybreaks[4]

\begin{document}

\columnsep 0.241in
\topskip 0.19in

\begin{CJK}{UTF8}{gbsn}
\title{A DRL-Empowered Multi-Level Jamming Approach for Secure Semantic Communication}

\author{\IEEEauthorblockN{Weixuan Chen and Qianqian Yang\textsuperscript{\textsection}
% , Yiping Duan$^{\dag}$, Xiaoming Tao$^{\dag}$, and Zhiguo Shi$^{*}$
}
% \author{\IEEEauthorblockN{Weixuan Chen$^{*}$ \thanks{
% This work is partly supported by NSFC under grant No. 62293481, No. 62201505, 
% partly by the National Key R\&D Program of China under Grant 2024YFE0200802.
% }, Qianqian Yang$^{*}$\textsuperscript{\textsection}, Shuo Shao$^{\dag}$, Zhiguo Shi$^{*}$, Jiming Chen$^{\ddag}$, Xuemin (Sherman) Shen$^{**}$}

\IEEEauthorblockA{{
% The State Key Laboratory of Industrial Control Technology
}
{College of Information Science and Electronic Engineering, Zhejiang University, Hangzhou, China}\\
% {$^{\dag}$ Department of Electronic Engineering, Tsinghua University, Beijing, China}\\
% {$^{\dag}$ Department of System Science, University of Shanghai for Science and Technology, Shanghai, China}\\
% {College of Information Science and Electronic Engineering, Zhejiang University, Hangzhou 310007, China}\\
% {College of Information Science and Electronic Engineering, Zhejiang University, Hangzhou 310007, China}\\
% Hangzhou, China \\
\{weixuanchen, \textsuperscript{\textsection}qianqianyang20\}@zju.edu.cn}
% \{weixuanchen, \textsuperscript{\textsection}qianqianyang20, shizg\}@zju.edu.cn, \{yipingduan, taoxm\}@mail.tsinghua.edu.cn}

\thanks{This work is partly supported by the NSFC under grant No. 62293481, No. 62571487, No. 62201505, by the National Key R\&D Program of China under Grant 2024YFE0200802, and by the Zhejiang Provincial Natural Science Foundation of China under Grant No. LZ25F010001.  (Corresponding author: Qianqian Yang.)}

}

% shuoshao@usst.edu.cn
\maketitle

\begin{abstract}

Semantic communication (SemCom) aims to transmit only task-relevant information, thereby improving communication efficiency but also exposing semantic information to potential eavesdropping. 
In this paper, we propose a deep reinforcement learning (DRL)-empowered multi-level jamming approach to enhance the security of SemCom systems over MIMO fading wiretap channels.
This approach combines semantic layer jamming, achieved by encoding task-irrelevant text, and physical layer jamming, achieved by encoding random Gaussian noise.
These two-level jamming signals are superposed with task-relevant semantic information to protect the transmitted semantics from eavesdropping. 
A deep deterministic policy gradient (DDPG) algorithm is further introduced to dynamically design and optimize the precoding matrices for both task-relevant semantic information and multi-level jamming signals, aiming to enhance the legitimate user's image reconstruction while degrading the eavesdropper's performance.
To jointly train the SemCom model and the DDPG agent, we propose an alternating optimization strategy where the two modules are updated iteratively. 
Experimental results demonstrate that, compared with both the encryption-based (ESCS) and encoded jammer-based (EJ) benchmarks, our method achieves comparable security while improving the legitimate user's peak signal-to-noise ratio (PSNR) by up to approximately 0.6 dB.

\end{abstract}
%Unlike existing methods that rely on key-sharing-based encryption schemes, shared prior knowledge, or channel advantages, our framework addresses a more practical and challenging scenario where the legitimate user and the eavesdropper experience comparable channel conditions without relying on such assumptions.


\section{Introduction}

% 第一段

% 无线数据流量的大幅度增长和XXX的XXX（自行补充）是6G时代的无线通信所面临的重大挑战。传统的通信技术专注于每个符号的准确恢复，而未能关注到传输数据所包含的意义，从而会浪费无线通信的资源。近年来，语义通信作为一种新颖的面向任务的通信范式，显著提高了通信的效率和鲁棒性，受到了广泛的关注。语义通信专注于提取并传输与语义任务相关的有用信息，而非全部的内容，从而去除了冗余信息，减少了传输数据量与通信延迟，有望成为6G网络中的一项核心技术。语义通信的一项代表性和开创性的技术为deep joint source-channel coding (DeepJSCC)。它利用深度神经网络从源数据中提取和重建有用的语义信息，并显著优于传统的数字通信系统。在这项工作的基础上，许多语义通信研究在各种模态的数据传输上都取得了优异的进步。

% The rapid growth of wireless data traffic, along with increasing demands for low latency and high reliability, poses significant challenges for communication systems in the upcoming 6G era. 
% 
% Traditional methods prioritize bit-level accuracy, which often leads to inefficient use of wireless resources.
% 
% 
% One of the pioneering technologies of SemCom is \textit{deep joint source-channel coding} (DeepJSCC) \cite{bourtsoulatze2019deep}, which leverages deep neural networks (DNNs) to extract and reconstruct semantically significant features from source data.
% 
% Building on this foundation, 

% Semantic communication (SemCom) has emerged in recent years as a promising paradigm that transmits task-relevant information rather than raw symbols, thereby enhancing communication efficiency.

% 
% Although SemCom has achieved remarkable performance, its inherent design, which focuses on transmitting only important semantics while removing much of the natural redundancy, makes the transmitted data especially vulnerable to eavesdropping, as intercepted signals may directly reveal sensitive content.

Semantic communication (SemCom) has emerged as a promising paradigm that improves communication efficiency by transmitting only task-relevant information, rather than raw symbols.
Significant progress has been made in various data modalities, including text \cite{mao2024gan}, image \cite{chen2023deep}, speech \cite{chen2024perceptually}, and video \cite{gao2024semantic}.
However, by focusing primarily on the essential semantics, SemCom typically eliminates traditional channel coding, which adds redundancy to improve reliability and provide implicit security.
The absence of this redundancy introduces a vulnerability: transmitted data becomes more susceptible to eavesdropping, as intercepted signals can directly reveal sensitive content.
As a result, secure SemCom over wiretap channels has become a crucial research direction and has recently attracted considerable attention \cite{chen2025knowledge}.
Existing approaches can be broadly categorized into three groups: 
adversarial training-based methods, encryption-based methods, and physical layer-based methods.

% As a result, secure SemCom has become an important research direction \cite{chen2025knowledge}, with recent studies exploring its design over wiretap channels.

% 第二段： 
% 尽管语义通信取得了优异的性能，由于语义通信仅传输与任务相关的重要信息，这会带来严重的安全问题。无线信道的开放性和传输数据的语义重要性使得语义通信容易遭受窃听攻击，从而导致源数据的敏感信息的泄露。这促使了在窃听信道上的安全语义通信作为一个关键研究领域的兴起。（接下来介绍我给你发的这几篇参考文献，要指出它们做了什么，以及总结他们的缺点。可以把他们分类进行介绍。注意开头要说XX \textit{et al.} 提出的XX）

% 
% 
% Chen \textit{et al.} introduced a differential private (DP)-based SemCom scheme for image transmission that disentangles private and public features and adds learnable noise to preserve privacy under equal-SNR wiretap scenarios. 
% 
% On the cryptographic front, Tung \textit{et al.} designed DeepJSCEC, integrating cryptographic encryption into DeepJSCC to defend against chosen-plaintext attacks, independent of the eavesdropper's channel knowledge. 
% 

From the adversarial training perspective,
Erdemir \textit{et al.} \cite{erdemir2022privacy} proposed a variational autoencoder (VAE)-based deep joint source-channel coding (DeepJSCC) scheme that enhances security by maximizing the mutual information (MI) between the source and the legitimate channel output while penalizing information leakage.
However, its security relies on variational approximations of MI, which may reduce its robustness against stronger adversaries.
From the key-sharing-based encryption perspective, Luo \textit{et al.} \cite{luo2023encrypted} developed an encrypted SemCom system (ESCS) that integrates a semantics-based symmetric cryptosystem learned through neural networks (NNs), and further proposed an adversarial encryption training strategy.
However, it requires a secure channel for key exchange, which remains a limitation for practical deployment.
More recent efforts have explored the use of homomorphic encryption in SemCom.
Meng \textit{et al.} \cite{meng2025secure} showed that semantic features can be retained in homomorphically encrypted data and proposed a privacy-preserved DeepJSCC framework, where activation and pooling functions are modified to support homomorphic operations.
% 
% explored homomorphic encryption for SemCom, enabling semantic feature extraction directly from ciphertext without decryption. Their method preserves semantic integrity while enhancing privacy in image-based tasks. 
% However, both approaches rely on the availability of a secure channel for key exchange, which remains a limitation for deployment in real-world scenarios.
% 

From the physical layer perspective, 
recent efforts adopt Shannon capacity and secrecy capacity as fundamental design principles.
Li \textit{et al.} \cite{li2024secure} proposed a two-phase training framework that balances reliable semantic recovery and reduced information leakage.
The training process employs loss functions derived from Shannon and secrecy capacities, while a secure BLEU metric is used to evaluate semantic security against eavesdropping.
% 
% , a DNN-based secure SemCom system with two-phase training: one phase for reliable recovery and another for minimizing information leakage, using capacity-guided loss functions and the S-BLEU metric for evaluation.
% 
% 
Beyond capacity-based designs, other works have explored the use of artificial noise to enhance security.
He \textit{et al.} \cite{he2025diffusion} proposed a framework where artificial noise is added at the transmitter and removed using denoising diffusion models at the legitimate receiver, thereby achieving a trade-off among communication reliability, privacy protection, and covertness.
However, the use of generative denoising introduces considerable computational complexity and latency.
Considering explicitly controllable system security, Chen \textit{et al.} \cite{chen2024nearly} superposed semantic information and random data on layered constellations under a tunable power allocation coefficient (PAC), thereby ensuring reliable decoding for the legitimate user and near-random decoding for the eavesdropper.
However, the security of this method depends on a sufficient channel signal-to-noise ratio (SNR) gap between the legitimate user and the eavesdropper.

To address the limitations of existing approaches,
we propose a novel deep reinforcement learning (DRL) \cite{arulkumaran2017deep}-empowered multi-level jamming framework for secure SemCom over MIMO fading wiretap channels. 
The main contributions of this paper are as follows.
% tailored for the challenging comparable-condition MIMO fading wiretap channel scenario.
% % 
% the main contributions of this paper are as follows.
% 
\begin{itemize}
% 
% 
% \item \textit{Multi-level jamming mechanism}:
\item
We propose a novel secure SemCom framework that innovatively introduces a multi-level jamming mechanism. 
Unlike existing methods that typically provide protection at only a single level, our framework simultaneously incorporates semantic layer and physical layer jamming signals to protect the semantic information, offering more comprehensive and robust privacy protection.
\item
% \item \textit{DRL-empowered precoding and transmission}:
We pioneer the integration of DRL into secure SemCom systems, which, to the best of our knowledge, has not been thoroughly explored in existing studies.
Considering the complexity and challenges of the channel setting, we employ DRL to optimize the design of precoding matrices. 
These matrices empower the effective superposition and transmission of semantic information and jamming signals, thereby enhancing both task performance and system security.
\item
% \item \textit{Practical secure SemCom framework}:
% 
Our framework addresses a highly challenging MIMO fading wiretap channel scenario, where both users experience comparable channel conditions, and Eve is assumed to have access to Bob's receiver architecture and training dataset.
Furthermore, unlike existing methods, our approach does not rely on secret key exchange, shared prior knowledge, or channel advantage, thereby reducing communication overhead and enhancing practicality.
Moreover, the DRL agent introduces no additional computational overhead during inference.
% is trained alternately with the SemCom model and 
% introduces no additional computational complexity during inference.
% 
% \item \textit{Key-exchange-free and inference-efficient design}:
% % The proposed multi-level jamming approach does not rely on key-based encryption. This removes the need for secure key exchange channels, thus reduce the communication overhead.
% % In addition, the DRL empowered multi-level jamming approach does not introduce additional computational complexity during the inference stage.
% It is worth noting that the proposed framework does not rely on key-based encryption, thereby eliminating the need for secure key exchange channels and reducing communication overhead. 
% Moreover, the DRL empowered multi-level jamming scheme introduces no additional computational complexity during the inference stage, thus making it well-suited for real-time secure communication applications.
% 
\item
We conduct extensive experiments to demonstrate that, compared with the key-sharing-based encrypted SemCom benchmark (ESCS) \cite{luo2023encrypted} and the encoded jammer-based benchmark (EJ) \cite{xu2024coding}, our approach achieves comparable security levels while improving the peak signal-to-noise ratio (PSNR) performance of the legitimate user by up to approximately 0.6 dB.
\end{itemize}

\section{System Model}

\begin{figure*}[t]
\begin{center}
\centerline{\includegraphics[width=1\linewidth]{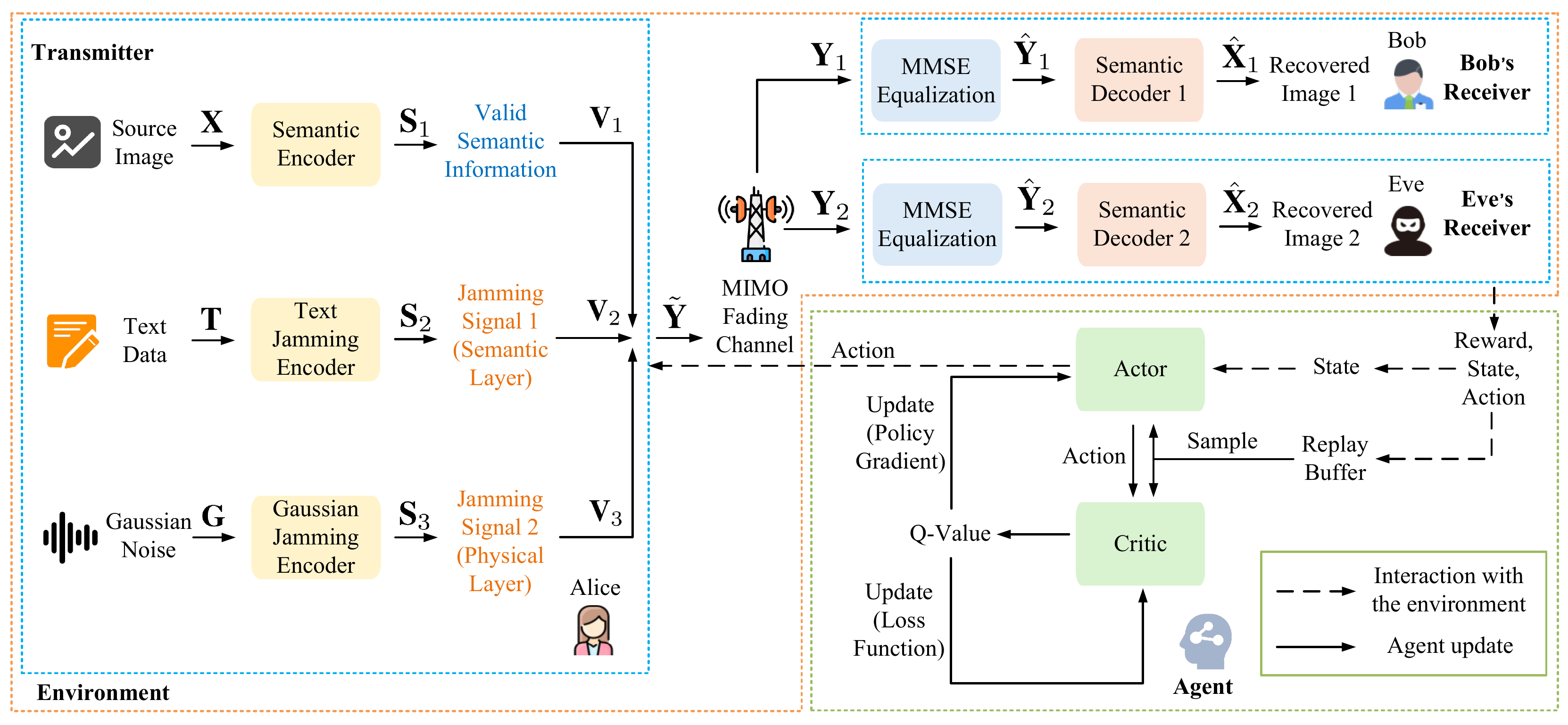}}
\caption{The framework of our proposed secure SemCom system.}
\label{fig.framework}
\end{center}
\vskip -0.4in
\end{figure*}

% 在本文中，我们考虑一个SemCom系统，被设计用于over MIMO窃听fading信道进行安全的图像传输，如Fig.X所示。该系统包括三个实体：一个transmitter(Alice)，一个合法receiver也被称为合法user(Bob)，以及一个窃听者(Eve)。Alice的目标是通过MIMOfading信道向Bob发送一张私有的源图像X，使得Bob能够很好地重建源图像X，同时不希望其他人能够重建它。Eve则通过一个和Bob的信道条件相同的MIMOfading信道被动地截获Alice传输的语义信息，并试图基于该语义信息重建源图像X。本文考虑了Bob和Eve的信道条件相同的现实和具有挑战性的场景，在这个场景下更能说明本文所提出的secureSemCom方法的有效性。所考虑的系统的总体目标就是使得Bob能够准确地恢复X，同时最大限度地减少对Eve的信息泄露。接下来，我们将详细地介绍所考虑的系统的模型。Section IV将提供所提出的secureSemCom方法的技术细节。

In this paper, we consider a SemCom system designed for secure image transmission over MIMO fading wiretap channels, as illustrated in Fig.~\ref{fig.framework}. 
We consider a MIMO SemCom system with $N_m \times N_n$ antennas.
The system involves three entities: a transmitter (Alice), a legitimate receiver (Bob), and an eavesdropper (Eve). 
Alice aims to transmit a private source image $\textbf{X}$ to Bob over MIMO fading channels so that Bob can accurately reconstruct the image while effectively preventing any unauthorized users from recovering it.  
Eve passively intercepts the semantic information through comparable MIMO fading channels and attempts to reconstruct $\textbf{X}$. 

In addition to the source image $\textbf{X} \in \mathbb{R}^{H \times W \times C}$, other information is also included as input to the transmitter. 
We denote the entire input as $\textbf{Z}$. Alice employs a transmitter to encode the input $\textbf{Z}$ in order to protect $\textbf{X}$, generating the semantic information to be transmitted, denoted as $\textbf{Y} \in \mathbb{R}^{N_{m} \times L_c}$:
\begin{equation}
    \textbf{Y} = f_{\mathrm{Alice}}\left( \textbf{Z} \right),
\end{equation}
where $f_{\mathrm{Alice}}$ represents the transmitter.

% 接下来，Y经过平均功率归一化来满足功率约束P，得到Y~，Y~随后被输入到一个MIMO Rayleigh fading channel中，以传输给Bob。Eve经过他自己的MIMO Rayleigh fading channel截获传输的语义信息。在本文中，由于我们假设窃听者和合法用户的信道条件是一致的，因此Bob和Eve接收到的语义信息分别为Y1和Y2，表示为：

% Subsequently, $\textbf{Y}$ is normalized to satisfy the average transmit power constraint $P$, resulting in the normalized signal $\tilde{\textbf{Y}}$.

Subsequently, $\textbf{Y}$ is normalized with respect to its average power in order to satisfy the transmit power constraint $P$ and then converted into a complex-valued representation, resulting in the normalized signal $\tilde{\textbf{Y}} \in \mathbb{C}^{N_{m} \times \frac{L_c}{2}}$.
This signal $\tilde{\textbf{Y}}$ is then transmitted through MIMO Rayleigh fading channels to Bob.
Meanwhile, Eve intercepts the transmitted semantic information through her own channels.
% 
% Before transmission, the signal is converted into a complex form and converted back into a real signal at the receiver.
% 
% In this paper, we assume that the channel conditions for both the legitimate user and the eavesdropper are comparable.
% 
Accordingly, the semantic information received by Bob and Eve are denoted as
${\textbf{Y}}_1 \in \mathbb{C}^{N_{n} \times \frac{L_c}{2}}$ and
${\textbf{Y}}_2 \in \mathbb{C}^{N_{n} \times \frac{L_c}{2}}$, respectively, and are expressed as
\begin{equation}
    {\textbf{Y}}_{1/2} = \textbf{H}\tilde{\textbf{Y}} + \textbf{N}_{1/2},
\end{equation}
% 
% 其中\textbf{H}表示信道矩阵between Alice和Bob/Eve。
where $\textbf{H} \in \mathbb{C}^{N_{n} \times N_m}$ denotes the channel matrix between Alice and Bob/Eve. 
Considering a conventional setting, we assume $N_m = N_n$.
The channel coefficients $\textbf{h}$ follow the distribution $\textbf{h} \sim \mathcal{CN}(0, \textbf{I}_{N_n})$,
where $\textbf{I}_{N_n}$ denotes the $N_n \times N_n$ identity matrix.
$\textbf{N}_{1/2} \in \mathbb{C}^{N_n \times \frac{L_c}{2}}$ represents the additive white Gaussian noise (AWGN) term for Bob/Eve,
where each element follows a complex Gaussian distribution $\mathcal{CN}(0, \sigma_{1/2}^2)$.
The channel SNR between Alice and Bob/Eve can be calculated by 
% \begin{equation}
$\mathrm{SNR}_{\rm leg/\rm eve} = 10 \log_{10}\left(\frac{P}{\sigma_{1/2}^2}\right) \, (\mathrm{dB})$.
% \end{equation}

% 在接收端，Bob和Eve的目标都是尽可能准确地重建源图像X。Bob和Eve在将接收的语义信息Y1和Y2输入到它们的语义解码器之前，会先通过minimum mean-squared error(MMSE)均衡对接收的语义信息进行信号检测。MMSE均衡是一种常用的MIMO检测技术，可以在接收端减轻信道干扰。MMSE均衡是基于估计的CSI H进行的。在本文中，对于所提出的方法和benchmarks，估计的CSI都被假设为是perfect的，并且不考虑导频开销。MMSE均衡的过程如下：

At the receiver, 
before feeding the received semantic information ${\textbf{Y}}_1$ and ${\textbf{Y}}_2$ into their respective semantic decoders, Bob and Eve first perform signal detection on the received signals using minimum mean-squared error (MMSE) equalization \cite{yang2015fifty}.
MMSE equalization operates based on the estimated channel state information (CSI) $\hat{\textbf{H}} \in \mathbb{C}^{N_{n} \times N_m}$.
In this paper, the estimated CSI is assumed to be perfect.
The MMSE equalization process is as follows:
\begin{equation}
\bar{\textbf{Y}}_{1/2} = {\hat{\textbf{H}}}^{H}\left( \hat{\textbf{H}}{\hat{\textbf{H}}}^{H} + \frac{\sigma_{1/2}^{2}}{P}\textbf{I} \right)^{- 1}\textbf{Y}_{1/2}.
\end{equation}
Here, $\bar{\textbf{Y}}_1$ and $\bar{\textbf{Y}}_2$ denote the MMSE-equalized versions of ${\textbf{Y}}_1$ and ${\textbf{Y}}_2$, respectively.
$\sigma_{1/2}^{2}$ denotes the variance of the AWGN, and $P$ denotes the transmit power constraint.
Subsequently, $\bar{\textbf{Y}}_1$ and $\bar{\textbf{Y}}_2$ are converted back to real-valued representations, denoted by $\hat{\textbf{Y}}_1$ and $\hat{\textbf{Y}}_2$.

% After that, for Bob, 
% 在那之后，对于Bob，他使用他自己的接收机基于Y1hat重建源图像，得到X1。这个过程可以表示为：

After that, Bob uses his own receiver to reconstruct the source image based on $\hat{\textbf{Y}}_1$, resulting in $\hat{\textbf{X}}_1$:
\begin{equation}
    \hat{\textbf{X}}_1 = f_{\mathrm{Bob}}\left( \hat{\textbf{Y}}_1 \right),
\end{equation}
where $f_{\mathrm{Bob}}$ denotes Bob's receiver, and $\hat{\textbf{X}}_1$ denotes the recovered image of Bob.
Similarly, Eve uses her own receiver to reconstruct the source image based on $\hat{\textbf{Y}}_2$, resulting in $\hat{\textbf{X}}_2$:
\begin{equation}
    \hat{\textbf{X}}_2 = f_{\mathrm{Eve}}\left( \hat{\textbf{Y}}_2 \right),
\end{equation}
where $f_{\mathrm{Eve}}$ denotes Eve's receiver, and $\hat{\textbf{X}}_2$ denotes the recovered image of Eve.

%我们的研究目标是开发一个安全的SemCom框架，从而能够同时实现下列两个目标：（1）合法用户可以准确地恢复源图像，and（2）有效地限制窃听者能够获取的信息量。为了评估我们提出的安全的SemCom框架在满足这些目标方面的有效性，我们采用PSNR指标来评估合法用户和窃听者的图像重建质量。合法用户的PSNR性能越高，则说明合法用户更加准确地重建了源图像。窃听者的PSNR性能越低，则说明我们提出的安全SemCom框架的安全性越高，信息泄露越小。具体来说，PSNR指标被定义为（给出equation,10log_10*MAX^2/MSE (dB)）：

% Our research objective is to develop a secure SemCom framework that simultaneously achieves the following two goals:
% (1) the legitimate user can accurately reconstruct the source image, and
% (2) the amount of information obtained by the eavesdropper is effectively limited.
% % 
% To evaluate the effectiveness of our proposed secure SemCom framework in achieving these objectives, we adopt the PSNR metric to assess the image reconstruction quality for both the legitimate user and the eavesdropper. A higher PSNR for the legitimate user indicates more accurate reconstruction of the source image, while a lower PSNR for the eavesdropper implies stronger security of the proposed framework and less information leakage.
% 
% Specifically, the PSNR metric is defined as:
% \begin{equation}
% \text{PSNR} = 10 \log_{10} \left( \frac{\text{MAX}^2}{\text{MSE}} \right) (\mathrm{dB}),
% \end{equation}
% where $\text{MSE}$ denotes the mean squared error, and $\text{MAX}$ represents the maximum possible pixel value of the image.

\section{Proposed Method}

% \subsection{System Overview}

% 在本subsection中，我们详细陈述提出的secure semcom系统中的各个组成部分中的内部运算。提出的系统的组成部分包括transmitter，Bob和Eve的receiver，以及一个用于赋能transmitter的编码过程的DDPG agent。

% In this subsection, we provide a detailed description of the internal operations and functions of each component in the proposed secure SemCom system. These components include the transmitter, the receivers at Bob and Eve, and a DDPG agent that empowers the transmitter's encoding process.

\subsection{Transmitter}

%transmitter的输入Z包含三个部分，其中一个是有意义的源图像X，另外两个分别为文本数据T和随机生成的高斯噪声G。在transmitter中，我们采用多层次的干扰来保护源图像X的传输。具体来说，文本数据T在语义上和X完全不相关，因此可以被编码以作为semantic layer层面的干扰。G则可以被编码以作为physical layer层面的干扰。这两个不同层面的干扰可以很好地保护源图像的语义信息的传输，减轻信息泄露。

% \subsubsection{Transmitter}

The input $\textbf{Z}$ to the transmitter consists of three components: the source image $\textbf{X}$, the text data $\textbf{T}$, and a randomly generated Gaussian noise matrix $\textbf{G} \in \mathbb{R}^{H \times W \times C}$.
Each element in $\textbf{G}$ follows a standard Gaussian distribution $\mathcal{N}(0, 1)$.
To protect the transmission of the source image $\textbf{X}$, a multi-level jamming scheme is employed at the transmitter.
Specifically, the text data $\textbf{T}$, which is semantically irrelevant to $\textbf{X}$, is encoded to serve as the jamming signal at the semantic layer. Meanwhile, the Gaussian noise $\textbf{G}$ is encoded to serve as the jamming signal at the physical layer.
These two different levels of jamming can effectively protect the transmission of $\textbf{X}$ and help reduce the risk of information leakage.

% 具体来说，在transmitter中，Alice首先编码源图像X，并生成语义信息Y1，表示为
Specifically, at the transmitter, Alice first encodes the source image $\textbf{X}$ to generate its semantic information $\textbf{S}_1$, which is formulated as
\begin{equation}
    \textbf{S}_1 = f_{\mathrm{SE}}\left( \textbf{X}; \bm{\theta}^{\mathrm{SE}} \right),
\end{equation}
where $f_{\mathrm{SE}}$ denotes the semantic encoder, and $\bm{\theta}^{\mathrm{SE}}$ represents its learnable parameters.

% Alice随后用一个文本干扰编码器编码文本数据T，并生成semantic layer上的干扰信号Y2，表示为
Subsequently, Alice encodes the text data $\textbf{T}$ using a text jamming encoder (TJE) to generate the jamming signal at the semantic layer, denoted by $\textbf{S}_2$, which is formulated as
\begin{equation}
    \textbf{S}_2 = f_{\mathrm{TJE}}\left( \textbf{T}; \bm{\theta}^{\mathrm{TJE}} \right),
\end{equation}
where $f_{\mathrm{TJE}}$ denotes the text jamming encoder, and $\bm{\theta}^{\mathrm{TJE}}$ represents its learnable parameters.

% 最后，Alice用一个高斯干扰编码器编码高斯噪声G，并生成物理层上的干扰信号Y3，表示为
Finally, Alice encodes the Gaussian noise $\textbf{G}$ using a Gaussian jamming encoder (GJE) to generate the jamming signal at the physical layer, denoted by $\textbf{S}_3$, which is formulated as
\begin{equation}
    \textbf{S}_3 = f_{\mathrm{GJE}}\left( \textbf{G}; \bm{\theta}^{\mathrm{GJE}} \right),
\end{equation}
where $f_{\mathrm{GJE}}$ denotes the Gaussian jamming encoder, and $\bm{\theta}^{\mathrm{GJE}}$ represents its learnable parameters.

% 在生成语义信息和多层次干扰信号后，Alice采用由DDPG agent为其产生的预编码矩阵V1、V2和V3，分别对Y1、Y2和Y3进行预编码，随后对它们进行叠加编码，从而得到要发送的语义信息Y。该过程可以表示如下：

After generating the semantic information and multi-level jamming signals, Alice applies the precoding matrices $\textbf{V}_1$, $\textbf{V}_2$, and $\textbf{V}_3$, which are produced by the deep deterministic policy gradient (DDPG) agent (to be discussed later), to precode the signals $\textbf{S}_1$, $\textbf{S}_2$, and $\textbf{S}_3$, respectively.
These precoded signals are then combined through superposition coding to form the transmitted semantic information, denoted as $\textbf{Y}$.
The process is described by the following equation:
\begin{equation}
 \textbf{Y} = \textbf{V}_1\textbf{S}_1+\textbf{V}_2\textbf{S}_2+\textbf{V}_3\textbf{S}_3,
\end{equation}
where $\textbf{V}_1$, $\textbf{V}_2$, and $\textbf{V}_3$ are each of dimension $N_m \times N_n$, and $\textbf{S}_1$, $\textbf{S}_2$, and $\textbf{S}_3$ are each of dimension $N_n \times L_c$. As a result, $\textbf{Y}$ has a dimension of $N_m \times L_c$.

% where $\textbf{V}_1$, $\textbf{V}_2$ and $\textbf{V}_3$ 的维度均为 $N_m \times N_n$, 而 $\textbf{S}_1$, $\textbf{S}_2$ and $\textbf{S}_3$ 的维度均为 $N_n \times L_c$。所以$\textbf{Y}$的维度是$N_m \times L_c$。

\subsection{Receivers at Bob and Eve}

% Bob和Eve各自的接收机中都包含一个语义解码器。在本文中，我们假设Eve窃取了Bob的语义解码器的网络架构，但其参数仍需要自行训练。因此，Bob和Eve的接收机的架构是一致的。
% 对于Bob，他将经过MMSE均衡后的接收信号Y1hat输入到其语义解码器中，并且获得一个重建图像X1。该过程表示为

% Both Bob's and Eve's receivers are equipped with a semantic decoder.
In this paper, we assume that Eve has obtained the network architecture of Bob's semantic decoder.
% , but he still needs to train the model parameters independently. 
% As a result, the receiver architectures of Bob and Eve are identical.
% 
For Bob, the MMSE-equalized received signal $\hat{\textbf{Y}}_1$ is input into his semantic decoder to reconstruct the image $\hat{\textbf{X}}_1$:
\begin{equation}
    \hat{\textbf{X}}_1 = f_{\mathrm{SD1}}\left( \hat{\textbf{Y}}_1; \bm{\theta}^{\mathrm{SD1}} \right),
\end{equation}
where $f_{\mathrm{SD1}}$ denotes Bob's semantic decoder, and $\bm{\theta}^{\mathrm{SD1}}$ represents its learnable parameters.

% 类似地，Eve将经过MMSE均衡后的接收信号Y2hat输入到其语义解码器中，并且获得一个重建图像X2。该过程表示为
Similarly, Eve inputs the MMSE-equalized received signal $\hat{\textbf{Y}}_2$ into her semantic decoder to reconstruct the image $\hat{\textbf{X}}_2$:
\begin{equation}
    \hat{\textbf{X}}_2 = f_{\mathrm{SD2}}\left( \hat{\textbf{Y}}_2; \bm{\theta}^{\mathrm{SD2}} \right),
\end{equation}
where $f_{\mathrm{SD2}}$ denotes Eve's semantic decoder, and $\bm{\theta}^{\mathrm{SD2}}$ represents its learnable parameters.

\subsection{The DDPG Agent}

% 考虑到DRL具有强大的决策能力并且已经在资源分配领域得到了十分广泛的应用，本文提出了一种DRLempowered多层次jamming方案，用于安全的semcom系统overMIMOwiretapchannels。具体来说，我们利用一个DDPG agent来帮助实现本文所考虑的目标，即
% (1) the legitimate user can accurately reconstruct the source image, and
% (2) the amount of information obtained by the eavesdropper is effectively limited.

% Considering the strong decision-making capabilities of DRL and its widespread applications in resource allocation, this paper proposes a DRL empowered multi-level jamming scheme for secure SemCom systems over MIMO fading wiretap channels. Specifically, we employ a DDPG agent to achieve the following objectives:
% (1) ensure that the legitimate user can accurately reconstruct the source image, and
% (2) effectively limit the amount of information acquired by the eavesdropper.

% 
% 我们考虑一个标准的DRL设置，其中包含一个DDPG agent，它在离散时隙中观察环境（即Fig.1中除了agent以外的通信模型）。
% 在整个secure semcom系统的训练过程中，DDPG agent与环境进行交互以获取状态，随后将其输入到action network来获取transmitter中的三个预编码矩阵。环境执行action policy后被训练一定的epoch数，随后agent将从更新的环境中获得反馈，且评估其action policy的价值，并优化agent当中的神经网络的参数。该过程将重复一定的时隙数，最后确定一个最终的action policy，最终满足本文所考虑的目标。

We consider a standard DRL setup that includes a DDPG agent, which observes the environment 
% 
% (i.e., the system model excluding the DDPG agent, as shown in Fig.~\ref{fig.framework}) 
% 
in discrete time slots. 
During the training stage of the proposed system, the DDPG agent interacts with the environment to obtain the current state, which is then fed into the action network to generate three precoding matrices for the transmitter. 
After the environment executes the action policy and undergoes training for a certain number of epochs, the DDPG agent receives feedback from the updated environment, evaluates the value of its action policy, and optimizes the parameters of its networks.
% 
% This process is repeated over multiple time slots until a final action policy is obtained, thereby fulfilling the objectives of our proposed system.
% 
% 如Fig.1所示，DDPG agent中包含一个actor network和一个critic network。除此以外，还有actor network和critic network各自的副本，称为目标actor network和目标critic network，它们从actor network和critic network中获取参数以软更新自己的参数。
As shown in Fig.~\ref{fig.framework}, the DDPG agent consists of an actor network and a critic network. In addition, there are target networks, namely the target actor network and the target critic network, which serve as copies of the actor and critic networks, respectively. These target networks are softly updated by getting parameters from their corresponding main networks.

%我们首先介绍整个安全semcom通信框架的state-action definitions。

We begin by introducing the state-action definitions for the proposed secure SemCom system.
\begin{itemize}
    \item % state。
    \textit{State:} At each decision time step $t$, the environment state $\textbf{s}_t \in \mathbb{R}^7$ consists of the following variables:
\begin{equation}
\begin{split}
\textbf{s}_t = \left[ {\rm SNR}^{\rm leg}_t, {\rm SNR}^{\rm eve}_t, CU, \right. \\
\left. {\rm PSNR}^{\rm leg}_{t-1}, {\rm PSNR}^{\rm eve}_{t-1}, \mathcal{L}_{t-1}, \frac{t}{T} \right],
\end{split}
\end{equation}
    where ${\rm SNR}^{\rm leg}_t$ and ${\rm SNR}^{\rm eve}_t$ denote the SNRs of the legitimate user and the eavesdropper, respectively.
    $CU$ controls the compression ratio (CR) of the system. 
    ${\rm PSNR}^{\rm leg}_{t-1}$ and ${\rm PSNR}^{\rm eve}_{t-1}$ indicate the image reconstruction quality (PSNR) from the previous time step for Bob and Eve, respectively. $\mathcal{L}_{t-1}$ denotes the previous loss, and $\frac{t}{T}$ represents the relative training progress.
    % \item % action
    \item \textit{Action:} At each decision time step $t$, the action $\textbf{a}_t \in \mathbb{R}^{3 \cdot N_m \cdot N_n}$ is defined as the output of the actor network. It represents the flattened form of three precoding matrices $\{\textbf{V}_1, \textbf{V}_2, \textbf{V}_3\}$, each of dimension $N_m \times N_n$. 
    The action is subsequently reshaped into a tensor of size $3 \times N_m \times N_n$ as:
\begin{equation}
\textbf{a}_t \Rightarrow \left[ \textbf{V}_1, \textbf{V}_2, \textbf{V}_3 \right], \textbf{V}_i \in \mathbb{R}^{N_m \times N_n}, \; i=1,2,3.
\end{equation}
    % \item % reward
    \item \textit{Reward:} The reward $r_t \in \mathbb{R}$ at each decision time step is designed to simultaneously encourage high image quality for the legitimate user while limiting information leakage to the eavesdropper. It is defined as:
\begin{equation}
r_t = \mathrm{PSNR}^{\mathrm{leg}}_t - \lambda_r \cdot \mathrm{PSNR}^{\mathrm{eve}}_t,
\label{eq17}
\end{equation}
where $\mathrm{PSNR}^{\mathrm{leg}}_t$ and $\mathrm{PSNR}^{\mathrm{eve}}_t$ denote the image reconstruction quality for Bob and Eve, respectively. 
$\lambda_r$ is a tunable parameter that balances the trade-off between task performance and system security.
\end{itemize}

Subsequently, we describe how the DDPG agent is optimized to design effective precoding strategies. 
We define the actor network $\mu_\theta$ and the critic network $Q_\phi$.
Both networks are composed of several fully connected (linear) layers.
% 它们都是由几个linear layers所组成的。
% 
To stabilize the training process, target networks $\mu_{\theta'}$ and $Q_{\phi'}$ are maintained as time-delayed copies of the actor and critic networks, respectively.
A replay buffer $\mathcal{D}$ is used to store past transition tuples $(\textbf{s}_t, \textbf{a}_t, r_t, \textbf{s}_{t+1}, d_t)$, where $d_t$ denotes a terminal signal. 
At each decision time step, exploration noise sampled from an Ornstein-Uhlenbeck (OU) process $\mathcal{N}_t$ is added to the actor's deterministic output to encourage temporally correlated exploration in continuous action space:
\begin{equation}
    \textbf{a}_t = \mu_\theta(\textbf{s}_t) + \mathcal{N}_t.
\end{equation}

The critic network is optimized by minimizing the mean squared Bellman error over a minibatch of $B$ experiences sampled from $\mathcal{D}$:
\begin{equation}
    \mathcal{L}_{\text{critic}} = \frac{1}{B} \sum_{i=1}^{B} \left( y_i - Q_\phi(\textbf{s}_i, \textbf{a}_i) \right)^2,
\end{equation}
where the temporal-difference target $y_i$ is computed using the target actor and critic networks as:
\begin{equation}
    y_i = r_i + \gamma \cdot Q_{\phi'}(\textbf{s}_{i+1}, \mu_{\theta'}(\textbf{s}_{i+1})) \cdot (1 - d_i),
\end{equation}
with $\gamma$ denoting the reward discount factor.

The actor network is updated by applying the deterministic policy gradient, which aims to maximize the expected critic value under the current policy:
\begin{equation}
    \nabla_\theta \mathcal{J} \approx \frac{1}{B} \sum_{i=1}^{B} \nabla_\theta \mu_\theta(\textbf{s}_i) \nabla_{\tilde{\textbf{a}}} Q_\phi(\textbf{s}_i, \tilde{\textbf{a}}) \big|_{\tilde{\textbf{a}} = \mu_\theta(\textbf{s}_i)}.
\end{equation}
Accordingly, the following loss function is minimized in practice:
\begin{equation}
    \mathcal{L}_{\text{actor}} = - \frac{1}{B} \sum_{i=1}^{B} Q_\phi(\textbf{s}_i, \mu_\theta(\textbf{s}_i)).
\end{equation}

To ensure stable learning, the parameters of the target networks are updated via soft updates after each gradient step.
% :
% \begin{align}
%     \theta' &\leftarrow \tau \theta + (1 - \tau)\theta', \\
%     \phi' &\leftarrow \tau \phi + (1 - \tau)\phi',
% \end{align}
% where $\tau \in (0,1)$ is a small positive coefficient that controls the update rate.

% 随后介绍整个DDPG agent是怎么更新/优化的，需要自己定义所需的符号，并且需要利用公式去陈述，要按照我的代码去陈述清楚，不要遗漏关键的内容。agent的network的具体架构就简单用一两句话提及就可以了。要提及的是actor和critic network及其copies的优化过程，还要提及replay memory等。

\subsection{Training Strategy}

% 在本文中，我们提出了一个五阶段的训练策略来训练我们提出的secure Semcom system。The DDPG agent仅在训练阶段更新策略，它在测试阶段则直接输出策略。合法用户和窃听者具有相同的语义解码器的网络架构，以及相同的信道条件。因此，合法用户和窃听者之间的性能差距主要来自于我们提出的DRLempowered multi-level jamming scheme的有效性。

In this paper, we propose a five-stage training strategy to train the proposed system.  
% The DDPG agent updates its policy only during the training stage and directly applies the learned policy during the testing stage.
% 
% Both the legitimate user and the eavesdropper adopt the same network architecture for their semantic decoders and operate under identical channel conditions. 
% Therefore, the performance gap between the legitimate user and the eavesdropper mainly depends on the effectiveness of our proposed DRL empowered multi-level jamming scheme.

\subsubsection{Stage 1}

In this stage, we focus on training the semantic encoder $f_{\text{SE}}$ and the semantic decoder of the legitimate user $f_{\text{SD1}}$, while keeping all other modules frozen.
The transmitted semantic information in this stage is given by $\textbf{Y} = \textbf{S}_1$.
% 
% 
% The system is optimized using the MSE loss between $\textbf{X}$ and Bob's reconstructed image $\hat{\textbf{X}}_1$:
The system is optimized using
\begin{equation}
\mathcal{L}_{1} = {\rm MSE}(\textbf{X}, \hat{\textbf{X}}_1).
\end{equation}

% 要讲述在这个阶段，哪些部分是被训练的，这个阶段有哪些独特假设或者设置，这个阶段的损失函数是怎样的，等等。

\subsubsection{Stage 2}

In this stage, multi-level jamming is introduced into the system, but the DDPG agent is not yet involved. 
% 补充：多层次干扰被应用，但the DDPG agent没有被应用。
All modules are activated except for the eavesdropper's decoder and the DDPG agent.
The transmitted semantic information in this stage is given by $\textbf{Y}=\textbf{S}_1+\textbf{S}_2+\textbf{S}_3$.
The loss function remains:
\begin{equation}
\mathcal{L}_{2} = {\rm MSE}(\textbf{X}, \hat{\textbf{X}}_1).
\end{equation}

\subsubsection{Stage 3}
In this stage, the DDPG agent remains inactive. 
The eavesdropper's decoder $f_{\text{SD2}}$ is trained independently, with all other modules kept frozen. 
% This stage simulates an adversary attempting to reconstruct the source image from intercepted signals under realistic attack conditions.
% 
The transmitted semantic information in this stage is given by $\textbf{Y}=\textbf{S}_1+\textbf{S}_2+\textbf{S}_3$.
The loss function used is:
\begin{equation}
\mathcal{L}_{3} = {\rm MSE}(\textbf{X}, \hat{\textbf{X}}_2),
\end{equation}
where $\hat{\textbf{X}}_2$ denotes Eve's reconstructed image. 

\subsubsection{Stage 4}
In this stage, the DDPG agent is introduced and jointly trained with the SemCom model (environment) through an alternating update scheme. 
The transmitted semantic information in this stage is given by
% \begin{equation}
$\textbf{Y} = \textbf{V}_1\textbf{S}_1+\textbf{V}_2\textbf{S}_2+\textbf{V}_3\textbf{S}_3$.
% \end{equation}
Specifically, the training alternates between two stages:
\begin{itemize}
  \item \textit{Agent update stage:} At the beginning of each $K$-epoch interval, the DDPG agent observes the current environment state and generates a new action $\textbf{a}_t$, corresponding to a set of precoding matrices $\{\textbf{V}_1, \textbf{V}_2, \textbf{V}_3\}$. This action remains fixed throughout the subsequent $K$ epochs of environment training. After receiving the reward and state transition at the end of the $K$ epochs, the agent stores the resulting experience tuple in its replay buffer. If the buffer contains enough samples, the agent updates its actor and critic networks via mini-batch gradient descent.
  \item \textit{Environment update stage:} For the next $K$ epochs, the SemCom model is trained using the fixed action provided by the agent. During this stage, all modules are updated except for the eavesdropper's decoder $f_{{\rm SD2}}$. After $K$ epochs, the environment reports the reward and updated state to the agent, which then uses this feedback to refine its policy.
\end{itemize}
This alternating training procedure continues until the predefined number of training epochs is completed. The loss function used is:
\begin{equation}
\mathcal{L}_{4} = {\rm MSE}(\textbf{X}, \hat{\textbf{X}}_1) - \lambda_r \cdot {\rm MSE}(\textbf{X}, \hat{\textbf{X}}_2),
\end{equation}
where $\lambda_r$ is the same as in equation (\ref{eq17}).

\subsubsection{Stage 5}

In this stage, 
% unlike in stage 4, the DDPG agent does not perform any online updates. 
the optimal policy obtained in stage 4, selected based on the highest reward, is saved and directly applied to guide both the training and testing of the final system.
% 
% 
% As in stage 4, the eavesdropper's decoder $f_{\rm SD2}$ remains frozen throughout this stage. 
The transmitted semantic information is generated in the same way as in stage 4, and the same loss function is used.

\section{Performance Evaluation}

\subsection{Experimental Settings}

\subsubsection{Datasets}

% 在本文中，我们使用CIFAR-10数据集用于源图像X，其中包括50000张训练图像和10000张测试图像。这些图像的维度均为32x32x3。我们使用the proceedings of the European Parliament数据集用于文本数据T，并且将其进一步处理为50000条训练数据和10000条测试数据。
% \subsubsection{Datasets}

% \subsubsection{Datasets}

We use the CIFAR-10 dataset for the source image $\textbf{X}$.
% , which contains 50,000 training images and 10,000 testing images, all with dimensions of $32 \times 32 \times 3$. 
 % 
% 
For the text data $\textbf{T}$, we utilize the proceedings of the European Parliament dataset \cite{papineni2002bleu}.
% , which we further process into 50,000 training samples and 10,000 testing samples.

\subsubsection{Training Settings}

We set $N_m=N_n=4$. 
Both users are assumed to experience the same channel SNR, given by $\mathrm{SNR}_{\mathrm{leg}} = \mathrm{SNR}_{\mathrm{eve}} \in \{0, 5, 10, 15, 20\}\,\mathrm{dB}$. 
The CR is controlled by the variable $CU$. When $CU = 1$, the corresponding CR is $1/96$. The CR takes values from the set $CR \in \{1/96, 2/96, 3/96, 4/96, 5/96\}$.

For the DDPG agent, $K=15$, the total number of decision time steps $T$ is set to 500, and the trade-off coefficient $\lambda_r$ is set to 1.0. The replay buffer size is set to 1000, and the minibatch size $B$ is 128.
The discount factor $\gamma$ is set to 0.99, and the soft update coefficient $\tau$ is set to $1 \times 10^{-3}$.
The learning rate of the actor network is $3 \times 10^{-4}$, and that of the critic network is also $3 \times 10^{-4}$.
The L2 weight decay is set to $1 \times 10^{-4}$.
% 
% During training, the agent and the environment are updated alternately at intervals of $K=15$ epochs. In each interval, the agent generates a fixed action every $K$ epochs and then updates its policy based on the observed state transitions and corresponding rewards.

% Our experiments are conducted on a single NVIDIA RTX A6000 GPU.
% The batch size is set to 256, and during training, we use the Adam optimizer.
% Our experiments are conducted on a single NVIDIA RTX A6000 GPU with a batch size of 256. During training, we employ the Adam optimizer. The learning rate in the first three training stages is set to $1 \times 10^{-3}$ and maintained for 100 epochs. In the fourth training stage, the learning rate is adjusted to $5 \times 10^{-4}$ and maintained for $500K$ iterations, which correspond to 7500 epochs. In the fifth training stage, the learning rate is set to $2 \times 10^{-4}$ and maintained for 200 epochs.

% Our experiments are conducted on a single NVIDIA RTX A6000 GPU with a batch size of 256. During training, we use the Adam optimizer.
% 
The learning rate in the first three training stages is set to $1 \times 10^{-3}$ and maintained for 100 epochs.
In the fourth training stage, the learning rate is set to $5 \times 10^{-4}$ and maintained for $T\cdot K$ epochs, which corresponds to 7500 epochs.
In the fifth training stage, the learning rate is set to $2 \times 10^{-4}$ and maintained for 200 epochs.

\subsubsection{The Benchmarks}

% In this paper, we employ two types of secure communication benchmarks for comparison with the proposed method. 
The first benchmark is the ESCS scheme \cite{luo2023encrypted}, which is a key-based encryption approach designed for secure SemCom systems over wiretap channels.
The second benchmark is the EJ scheme \cite{xu2024coding}, which is a jamming-based scheme developed for secure communication systems over MIMO wiretap channels.  
% Both benchmarks employ the classical singular value decomposition (SVD) precoding and MMSE equalization.
In both benchmarks, we employ classical singular value decomposition (SVD) precoding and MMSE equalization.
% 
% The loss functions and other fundamental settings of these benchmarks are kept consistent with those of our proposed method.
% The loss functions and other fundamental settings of these benchmarks are kept consistent with those of our proposed method.

\subsection{Performance Comparison of Different Approaches}

\subsubsection{Different Channel SNRs}

\begin{figure}[t]
\begin{center}
\centerline{\includegraphics[width=0.98\linewidth]{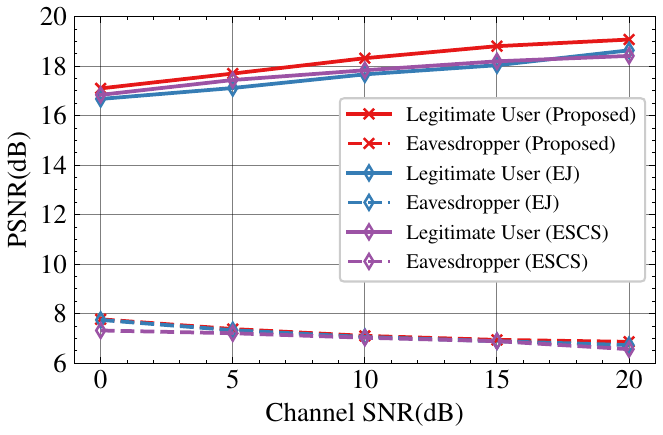}}
\caption{Comparison of PSNR performance for both users under different channel SNRs using different approaches. The CR is fixed at $1/96$, with $N_m$ and $N_n$ both set to 4. We set $\rm SNR_{leg} = \rm SNR_{eve}$.}
\label{fig.plot1_proc}
\end{center}
\vskip -0.3in
\end{figure}

In this subsection, we evaluate the PSNR performance of the proposed system under varying channel SNRs. 
% The comparison involves three approaches: our proposed method, the EJ benchmark \cite{xu2024coding}, and the ESCS benchmark \cite{luo2023encrypted}.
% 
The experimental results are shown in Fig.~\ref{fig.plot1_proc}, with a fixed CR of $1/96$.
% 
% We have $\mathrm{SNR}_{\mathrm{leg}} = \mathrm{SNR}_{\mathrm{eve}} \in \{0, 5, 10, 15, 20\}\,\mathrm{dB}$, with a fixed CR of 1/96, and both $N_m$ and $N_n$ set to 4.
% 
As shown in Fig.~\ref{fig.plot1_proc}, the PSNR of the legitimate user steadily improves as the channel SNR increases. Our proposed method consistently outperforms both EJ and ESCS across all channel SNRs, providing a notable performance gain due to its efficient DRL-empowered multi-level jamming strategy.
Meanwhile, the eavesdropper's PSNR remains at a consistently low level and exhibits only minor fluctuations with changes in channel SNR, indicating that the superposed semantic and physical layer jamming effectively prevents accurate image reconstruction for the eavesdropper.
While ESCS achieves slightly lower PSNR for the eavesdropper, our proposed method attains a more favorable trade-off by enhancing the legitimate user's reconstruction quality without significantly compromising security.
% 
% Overall, these results demonstrate that the proposed method achieves robust task performance while effectively mitigating information leakage to the eavesdropper under different channel SNRs, thus validating its superiority in balancing security and performance.

% \subsubsection{Different Numbers of Antennas}

\begin{figure}[t]
\begin{center}
\centerline{\includegraphics[width=0.98\linewidth]{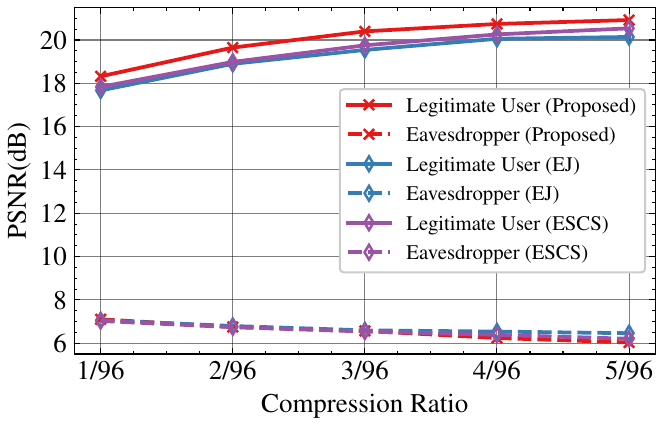}}
\caption{Comparison of PSNR performance for both users under different CRs using different approaches. We set $\rm SNR_{leg} = \rm SNR_{eve} = 10dB$, with $N_m$ and $N_n$ both set to 4.}
\label{fig.plot3_proc}
\end{center}
\vskip -0.3in
\end{figure}

\subsubsection{Different Compression Ratios}

In this subsection, we evaluate the PSNR performance of the proposed system under different CRs and compare it with the two benchmarks. 
% 
% 
% The evaluations are conducted with $\mathrm{SNR}_{\mathrm{leg}} = \mathrm{SNR}_{\mathrm{eve}} =10 \mathrm{dB}$, and both $N_m$ and $N_n$ set to 4.
% 
As shown in Fig.~\ref{fig.plot3_proc}, the PSNR of the legitimate user increases steadily as the CR grows. Among all evaluated methods, the proposed method consistently achieves the highest reconstruction quality across all CRs. Specifically, at a CR of $5/96$, our method attains a PSNR of 20.92 dB, outperforming EJ (20.14 dB) and ESCS (20.54 dB). 
In contrast to the legitimate user, the eavesdropper's PSNR exhibits a decreasing trend with higher CRs. At the highest CR, our method reduces the eavesdropper's PSNR to 6.03 dB, which is lower than that of EJ (6.46 dB) and ESCS (6.20 dB), indicating improved system security.
Overall, compared with the benchmarks, the proposed system consistently achieves a superior trade-off between task performance and system security, demonstrating high effectiveness and robustness under varying CRs.

\section{Conclusion}

In this paper, we proposed a SemCom framework that integrated a DRL-empowered multi-level jamming strategy to enhance transmission security over MIMO fading wiretap channels.
The framework was designed for a realistic and challenging scenario where both the legitimate user and the eavesdropper experienced comparable channel conditions, without relying on secret key exchange or any prior shared knowledge.
Specifically, two levels of jamming signals were introduced: semantic layer jamming, generated by encoding task-irrelevant text data, and physical layer jamming, generated by encoding random Gaussian noise.
These jamming signals were superposed with task-relevant semantic information to effectively obscure the transmitted semantics from the eavesdropper while maintaining high reconstruction quality for the legitimate user.
To optimize and empower this process, we employed the DDPG algorithm to design and adapt the precoding matrices for both task-relevant semantic information and multi-level jamming signals.
This approach effectively balanced the trade-off between the legitimate user's reconstruction quality and overall system security.
Moreover, the DDPG agent operated only during training, introducing no additional computational overhead during inference. 
Experimental results demonstrated that, compared with the ESCS and EJ benchmarks, our approach achieved comparable security while improving the legitimate user's PSNR by up to approximately 0.6 dB.
These results confirm that the proposed DRL-empowered multi-level jamming framework provides an effective and practical solution for secure SemCom systems operating under challenging conditions.

\bibliographystyle{IEEEtran}

\bibliography{myref}
% \end{thebibliography}

\vspace{12pt}
\end{CJK}
\end{document}